# Neutral Particle Analyzer for Studies of Fast Ion Population in Plasma

S. Polosatkin[a,b], V. Belykh[a], V. Davydenko[a,c], R.Clary[d], G. Fiksel[e], A. Ivanov[a,c], V. Kapitonov[a], D.Liu[e], V. Mishagin[a], M. Tiunov[a], and R. Voskoboynikov[a]

[a] Budker Institute of Nuclear Physics, 630090, Novosibirsk, Russia

[b] Novosibirsk State Technical University, 630092, Novosibirsk, Russia

[c] Novosibirsk State University, 630090, Novosibirsk, Russia

[d] Tri Alpha Energy inc., 92688, Ranco Santa Margarita CA, USA

[e] University of Wisconsine-Madison, 53706 Madison Wi, USA

**Abstract**

*Advanced neutral particles analyzer for plasma diagnostic with possibility of simultaneous measurements of energy distributions of D and H ions has developed in the Budker Institute of Nuclear Physics. The analyzer was used in two plasma facilities with injection of fast neutrals – on the MST reversed field pinch (University of Wisconsin) and the field reversed configuration C-2 (Tri Alpha Energy). In this paper, the design of the analyzer, calculation of efficiency of registration, results of analyzer calibration and experimental results from MST and C-2 experiments are presented.*

Nuclear Fusion, Plasma diagnostics, charge-exchange neutrals

## 1. Introduction

Analysis of energy distribution of charge-exchange neutrals is an informative tool for fusion plasma diagnostics. Both passive and active (beam-assisted) diagnostics of charge-exchange particles are widely used on most plasma facilities; different types of neutral particles analyzers based on using electric or magnetic fields for particles separation have been designed [1-5]. At the same time, resent achievements in electronics and particle detection techniques allow to design new type the analyzer with advanced performance and additional possibilities for plasma diagnostics.

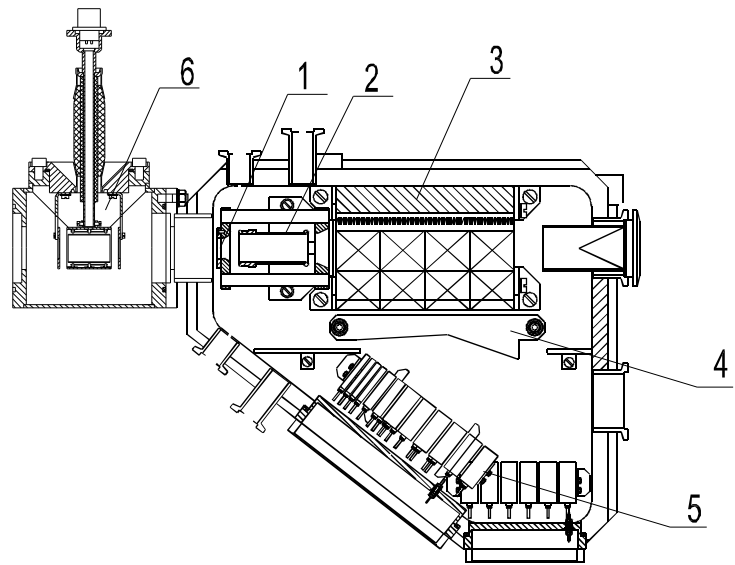

*Fig.1 Schematic of the ANPA; **1** – stripping foil, **2** – electrostatic lens, **3** – magnet-separator, **4** – mass-separating capacitor, **5** – detectors, **6** – calibration ion source*

A new Advanced Neutral Particle Analyzer (ANPA) has developed by the Budker Institute of Nuclear Physics (BINP) for application in fusion plasma facilities with injection of fast neutrals. The analyzer provides measurements in a broad energy range - from hundreds of eV for measurements of bulk ion plasma temperature to tens of keV to study the confinement of fast ions, formation of high energy ion tails, etc. In addition, the analyzer is capable to separate hydrogen and deuterium atoms and includes a built-in ion source for in-situ calibration. Two similar analyzers has installed on reversed field pinch MST (Univ. of Wisconsin) and FRC facility C-2 (Tri Alpha energy inc.) and have used for study of fast ion distribution in plasma.

**2. Design of the analyzer**

For informative application for plasma diagnostics the designed analyzer should satisfy the follows requirements:
- Measured temperatures of bulk plasma 0,4 - 3 keV
- Measured energies of fast ions - up to 30 keV
- Energy resolution 10-20%
- Temporal resolution up to 10 μs
- Possibility of separation of hydrogen (protium) and deuterium ions

The general operation principle of the analyzer is conventional for such types of diagnostics. (see e.g.[4]). Energetic neutrals appear in plasma due to charge exchange (CX) of plasma ions with either background hydrogen atoms or fast neutrals from diagnostic or heating beams. The CX neutrals travel across the magnetic field and enter the entrance aperture of the neutral analyzer. In the analyzer neutrals are stripped to ions, separated by energies and masses, and acquired by a set of detectors.

A scheme of the analyzer is shown on Fig.1. Neutrals entering analyzer are converted to ions in the stripping unit, which includes a stripping target (10-nm-width carbon film) **1** and electrostatic lens **2**. A target is biased to +7 kV, therefore energies of incoming particles shifted to those energies. Electrostatic lens serves for collecting of ions scattered during passing through the foil.

Separation of ions by energies is implemented in magnetic field of C-shape permanent magnet-separator **3**. The maximal magnetic field in the magnet gap is 0.6 T. The pole shape is optimized for focusing of the ions in two dimensions (horizontal and vertical) onto the detector array. The ion trajectories in the magnet depend on ion velocity, so deflection of an ion in the magnet is proportional to $(m_i \cdot E)^{0.5}$, where $m_i$ and $E$ – the mass and the energy of the ion.

After the magnet, the ions pass through a mass-separating capacitor **4** with a transverse (with respect to the ion velocity) electric field. The electric field in the capacitor causes a vertical shift of the ions that is proportional to mass of the ion. Thus, after passing the magnet and the capacitor the ions are dispersed in two dimensions according to their energies and masses. Measuring of ion current in the given point allows one to determine the flux of neutrals of specified mass and energy.

Miniature secondary electron multipliers (SEM) Magnum5900 are used for ion detection. Up to 22 detectors can be mounted in the analyzer which allows to cover an energy range 0-30 keV with resolution about 10%. Switching off the bias voltage on the stripping foil allow to shift this regions even to higher energies.

**3. Particles detection**

Secondary electron amplifier MAGNUM 5900 was selected to use in the analyzer. The feature of this SEM is miniature dimensions and relatively high mean current (up to 20 µA). Thus this detector may operate both in count and current modes.

Characteristics of the SEM were studied during the analyzer preliminary testing. Ion flux from calibration ion source was forward to entrance of the SEM. Source flux intensity may be regulated in wide range by cathode current for comparison of count and current modes of the SEM. 500 MHz Tektronix oscilloscope was used for signal registration.

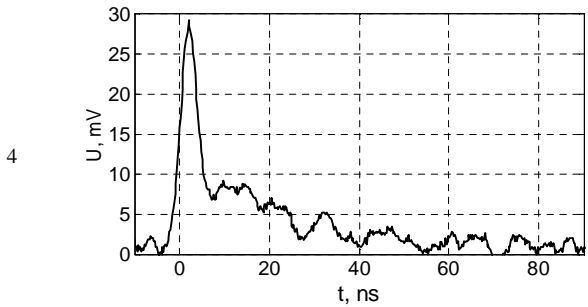

Fig.2 The waveform of single event pulse from SEM

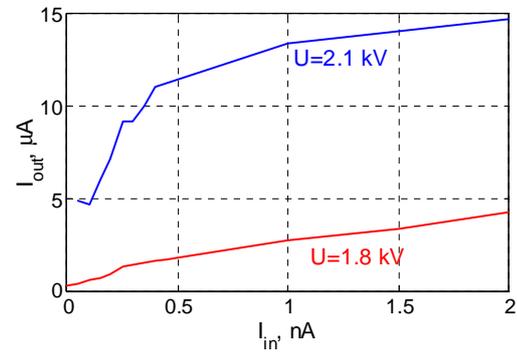

Fig.3 Dependence of SEM output current from input one

Single ion event pulse waveform is shown on Fig.2 for detector supply voltage 2.1 kV and load resistor 50 Ohm. The pulse duration on half width is 4.5 ns, integral charge (amplification) $1.5 \cdot 10^7$ electrons.

Dependence of output current from input one for supply voltages 1.8 and 2.1 kV is shown on Fig.3. For 2.1 kV supply saturation of output current at 10 µA clearly seen (the supplier claim 20 µA maximal current at 3 kV power supply). 300 V decrease of supply voltage cause four times increasing of linearity range of input current with corresponding drop of amplification. Thus the SEM may be exploited in the current mode but such regime requires careful control of detector linearity.

## 4. Analyzer calibration

The analyzer have equipped by miniature ion source **6** designated for in-situ calibration of the device. Operation principle of the ion source very similar to that used in high-vacuum ionization gauges (Bayard-Alpert scheme). It consists on tungsten filament emitting electrons and a grid with potential +200 V in respect to filament that serves as electrostatic trap. Both filament and grid are surrounded by screen electrode and biased to high voltage that accelerates ions generated by the source. Since the edges of the source are made from mesh and transparent for neutrals, the source can be mounted directly on the axis of the CX neutrals beamline that allow to produce in-situ calibration of the system.

Instrument functions for first six hydrogen channels measured with dismounted stripping foil are shown on fig.4a. Measured mean energies of channels are close to that predicted from ion trajectories calculations. Results of the same measurements with mounted foil are shown on fig.4b. The energy loss in the foil is in the range 1.7-2.2 keV for primary hydrogen ion energies 10-25 keV. This result is in some contradiction with known data for stopping power of protons in carbon that predict energy loss less then 1 keV [6]. Possible reasons of such difference are some surface effects that not taken into account in PSTAR database or increase of foil thickness due to contamination. If last assumption is correct regular energy calibration appear vital for correct measurements. In presented experiments maximal energy of ions from calibration source was restricted to 25 keV by HV power supply. Calibration of high-energy channels and study of stability of energy loss in the foil is now under way.

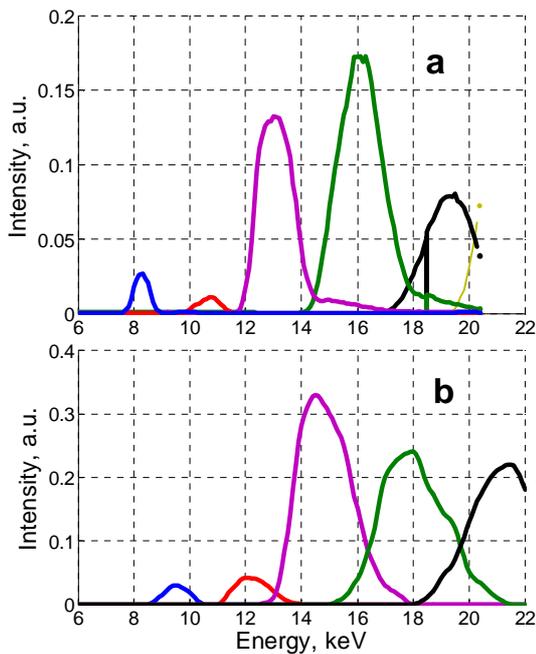

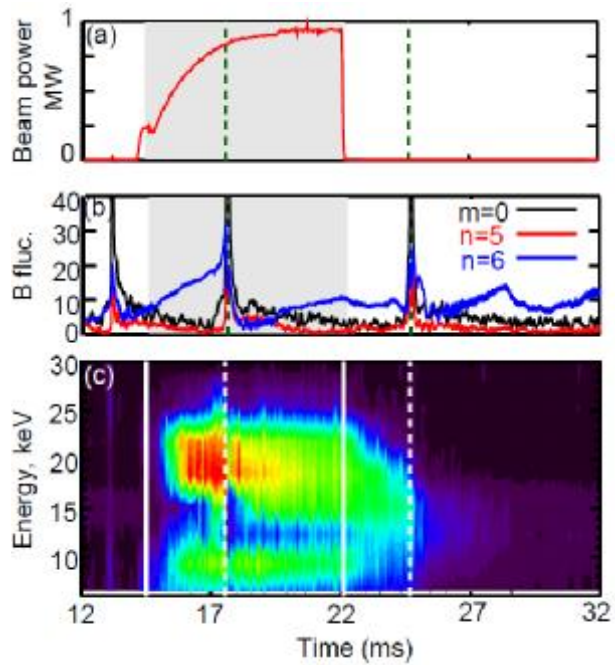

*Fig.4 Instrument functions of first five channels of registration, measured with (a) and without (b) stripping foil*

*Fig.5 Fast ion population dynamics on MST facility. **a** - neutral beam power; **b** magnetic fluctuation of m=0, and m=1, n=5,6 tearing modes; **c** - raw signals of ANPA in the 10 hydrogen channels. The two dashed lines mark the two time-slices with strong sawtooth.*

## 5. Study of fast ion population on MST facility

One analyzer was mounted on MST reversed field pinch [7] where it was used for studying of confinement of fast ions injected by powerful neutral beam injection (NBI) system. Dynamics of energy distribution of fast ion population is presented on Fig.5. Fast ions with energies up to $E_{inj}$=23.4 keV accumulates in plasma during neutral beam injection. After beam turn-off fast ions slowing down in plasma and high-energy edge of distribution function shifted to lower energies. The fast ions with energy lower than 10 keV are lost quickly and is not confined in plasma. Sawtooth events (dashed lines on the fig.5) cause strong redistribution of plasma and lead to partial or full loss of fast ion population. Those measurements of ion distribution function allow to estimate confinement time of fast ion in plasma that is in agreement with results of Monte-Carlo modeling of ion dynamics [8].

## 6. Conclusion

An Advanced Neutral Particle Analyzer was designed and fabricated in the Budker Institute of Nuclear Physics. The analyzer allows to perform a wide range of experiments to study the properties of hot plasma with population of fast ions. The analyzers have installed on two plasma facilities with injection of fast neutrals – on the MST reversed field pinch (University of Wisconsin) and the field reversed configuration C-2 (Tri Alpha Energy) where applied for study of hast ion population dynamics.

**Acknowledgments**

The work was done under financial support of Russian Ministry of Science and Education, programs "Scientific and educational staff of innovative Russia" and "National University of Science and Technology".